\documentclass[english]{IEEEtran}
\usepackage{graphicx}
\usepackage{hyperref}

\usepackage{amsmath}
\usepackage{float}

\begin{document}

\title{On the Impact of LiDAR Point Cloud Compression on Remote Semantic Segmentation}

\author{Tiago de S. Fernandes and Ricardo L. de Queiroz, 
\thanks{T. Fernandes was with the Electrical Eng. Dept. at Universidade de Brasilia, Brasilia, Brazil, e-mail: tiagotsf2000@gmail.com ; R. de Queiroz is with the Computer Science Dept. at Universidade de Brasilia, Brasilia, Brazil, e-mail: queiroz@ieee.org.}
}

\maketitle

\begin{abstract}
Autonomous vehicles rely on LiDAR sensors to generate 3D point clouds for accurate segmentation and object detection. 
In a context of a smart city framework, we would like to understand the effect that  transmission (compression) can have on remote (cloud) segmentation, instead of local processing.
In this short paper, we try to understand the impact of point cloud compression on semantic segmentation performance and to estimate the necessary bandwidth requirements. 
We developed a new (suitable) distortion metric to evaluate such an impact.  
Two of MPEG's compression algorithms (GPCC and L3C2) and two leading semantic segmentation algorithms (2DPASS and PVKD) were tested over the Semantic KITTI dataset. 
Results indicate that high segmentation quality requires communication throughput of approximately 0.6 MB/s for G-PCC and 2.8 MB/s for L3C2. 
These results are important in order to plan infrastructure resources for autonomous navigation.
\end{abstract}

\begin{IEEEkeywords}
lidar point cloud, compression, semantic segmentation, metrics
\end{IEEEkeywords}

\section{Introduction}

Autonomous vehicles rely on light detection and ranging (LiDAR) sensors to perceive their surroundings \cite{lidar_autonomous_vehicles}.
These sensors generate enormous amounts of data to create detailed 3D Point Clouds (PC), which are subject to semantic segmentation and object detection. 
In order to process, transmit or store this data, compression is essential \cite{pc_processing_autonomous_driving}.
International bodies such as the Moving Picture Experts Group\footnote{\url{www.mpeg.org}} (MPEG) are actively developing standard PC compressors \cite{8571288}.

An important processing step for autonomous systems is semantic segmentation, a key computer vision technique for interpreting point clouds. It assigns a semantic class to each point (e.g., car, person, road), enabling systems to understand their environment and to make critical real-time decisions. The integrity and quality of the data are crucial to ensure the accuracy of the semantic segmentation algorithms \cite{lidar_semantic_segmentation_survey}, which can reflect in the safety of the autonomous driving system.
Our objective is to evaluate the impact of point cloud compression on the performance of semantic segmentation algorithms. 

A key issue for such an evaluation is to properly select a distortion metric. Ideally, one would like to reduce or eliminate fatalities, but there are no settled algorithms relating segmentation quality to autonomous driving patterns.
Hence, we are interested in maintaining the quality of the segmentation algorithm, i.e. we want to minimize modifications into the segmented classes caused by the compression algorithms. 
Conventional point cloud distortion metrics imply point correspondence and are not suitable for many reasons.
For example, points can disappear or change class label, and objects may disappear, appear, or change their aspect. We also want to take into account if humans are involved in the class change. 

We use the popular Semantic KITTI dataset \cite{semantic_kitti} for tests.
Its website hosts a semantic segmentation competition\footnote{\url{http://www.semantic-kitti.org/tasks.html\#semseg}. Access was made in October 2023.}, from which we selected two top performers for tests: 
2D Projection-based Aggregated Segmentation Scheme (2DPASS) \cite{yan20222dpass} and the Point-Voxel Knowledge Distillation (PVKD)  \cite{pvkd}. 
We assume them to be state-of-the-art segmentation for our purposes and encourage the reader to look at the references for any details. 
In order to compress the point clouds, we use two MPEG compression algorithms: Geometry-based Point Cloud Compression (G-PCC) \cite{gpcc} and the Low-Latency Low-Complexity Codec (L3C2) \cite{l3c2}.
G-PCC is more sophisticated and we used octree coding of the geometry and RAHT \cite{Queiroz2016} for attributes.  
L3C2 is a simpler algorithm developed to address the specific needs of LiDAR point clouds in real-time 
energy-constrained applications. 
We want to find the maximum compression one can apply before substantially degrading segmentation quality.

\begin{figure}[t]
\centering
\includegraphics[width=.7\linewidth]{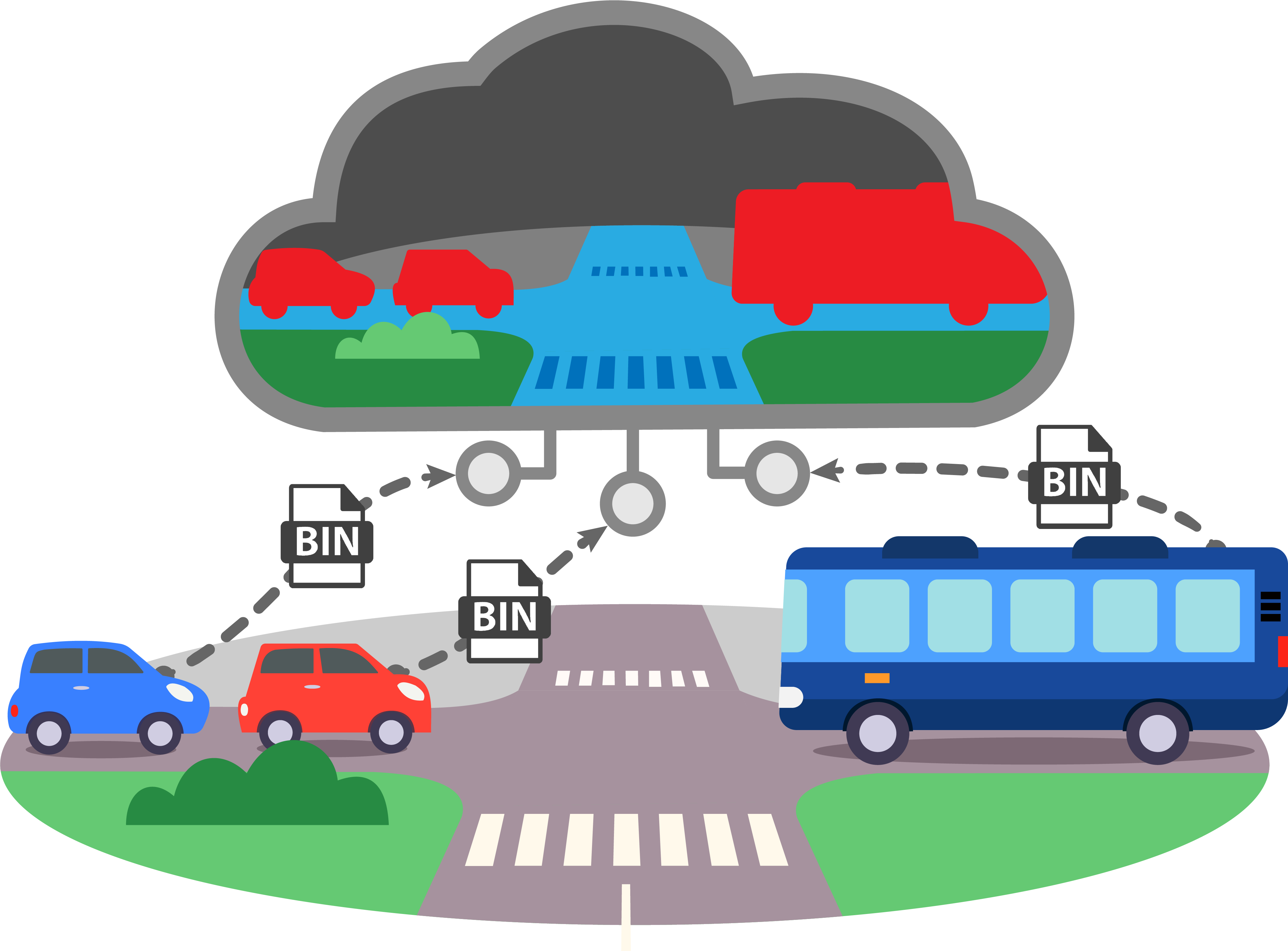}
\vspace{-1em}
\caption{Remote LiDAR Semantic Segmentation scenario.}
\label{fig:scene_tcc}
\end{figure}

The compression impact on PC segmentation has also been recently investigated \cite{literature_ref_1, literature_ref_2, literature_ref_3}. 
We, however, considered more up-to-date compressors and segmenters and we propose a new distortion metric.  
Results are still incipient for practical purposes and we believe all these analyses to be complementary. 

\section{Proposed Semantic Segmentation Metric}





Evaluating segmentation quality by comparing it to the ground truth is not trivial. This is so  because the compressed point cloud may be distorted and contain fewer points than the original, leading to the loss of direct label correspondence.  

Our approach provides a reliable way to compare labels despite these distortions and point reductions. For each point in the uncompressed point cloud, we identify the closest point in the compressed point cloud and compare their labels. The error metric is based on the proportion of mismatches for a given label. Importantly, the metric is only computed in the direction from the uncompressed to the compressed point cloud, in order to prevent the number of points in the compressed cloud from directly influencing the result. A segmentation could still perfectly represent all the information present in the original point cloud even if less points are available.

Given the uncompressed point cloud $X$, the compressed one $Y$, and a label $\lambda$, let $X_i$ be a point in $X$ and $Y_{Xi}$ the closest point to $X_i$ in $Y$ using Euclidean distance. 
Let $L(.)$ return the label of the given point.  
Define the set $S_\lambda$ as the set with all matching pairs involving label $\lambda$:
\begin{equation}
    S_\lambda = \{(X_i, Y_{Xi}) \ |\ (L(X_i) = \lambda) \text{ or } (L(Y_{Xi}) = \lambda) \} .
\end{equation}

We similarly define the subset of $S_\lambda$ wherein there is a label mismatch, i.e.: 

\begin{equation}
   E_\lambda = \{S_\lambda \ | \ L(X_i) \neq L(Y_{Xi})\} .
\end{equation}

Then, our metric is:

\begin{equation}
\delta(X, Y, \lambda) = \frac{|E_\lambda|}{|S_\lambda|} . 
\label{eq:metric}
\end{equation}

\noindent 
This metric is well defined even when an entire label is erased, and does not confuse geometry distortion with segmentation errors, since it does not use the distance between corresponding points. 
$\delta$ is an error that fits inside the range [0, 1], where $\delta=0$ means perfect (comparative) segmentation.

We further refine the metric potentially giving a higher weight for specific types of errors. 
For example, we believe that turning the human label ($\lambda_h$) into another label can be more critical than other errors. An $\alpha$ term is then introduced, adding a weight to the metric when a human label turns into a non-human one. 
If 

\begin{equation}
    E_h = \{ S_{\lambda_h} | L(X_i) = \lambda_h ; L(Y_{Xi}) \neq \lambda_h\}
\end{equation}

\noindent
is the set of matching pairs where human points become mislabeled after compression, the metric for human labels may be: 

\begin{equation}
 \delta(X, Y, \lambda_h) = 
    \frac{
        |E_{\lambda_h}| + \alpha \cdot | E_h | 
        }{
    |S_{\lambda_h}| + \alpha \cdot | E_h |  
    } .
\end{equation}

Note that for $\alpha = 0$ the metric is the same as for other labels. For other labels, other than human ($\lambda_h$), the metric is the one in (\ref{eq:metric}).







\section{Results}

In the Semantic KITTI dataset, each PC scan comes with 3-coordinates geometry, one attribute (reflectance), and a segmentation label. 
There are 28 class labels including {\em road}, {\em sidewalk}, {\em terrain}, {\em vegetation}, {\em human}, etc.
Six sequences (11, 12, 13, 14, 15, and 18) were arbitrarily chosen for tests, using 101 scans of each. 
We used 2DPASS and PVKD segmentation with their default settings. 


MPEG common test conditions \cite{ctc_ref} (CTC) provides six compression settings for each compressor, labeled from \textit{R01} to \textit{R06}, where \textit{R06} yields the best quality and least compression. 
The settings define a quantization parameter (QP) for compressing the transformed attributes, along with some general configurations. 
Furthermore, the settings affect geometry by directly quantizing the geometry positions with a Quantization Scale (QS). In essence, the geometry is multiplied by QS and rounded. In some cases, this is similar to pruning the octree.

\begin{table}[]
\caption{Average Throughput and Quantization Scale for each compression rate.}
\label{table:throughput_qs}
\centering
\begin{tabular}{c|cc|c}
\cline{2-3}
\multicolumn{1}{l|}{}                  & \multicolumn{2}{c|}{Throughput (MB/s)}             &   \\ \hline
\multicolumn{1}{|c|}{Compression} & \multicolumn{1}{c|}{\textbf{G-PCC}} & \multicolumn{1}{c|}{\textbf{L3C2}} & \multicolumn{1}{c|}{\textbf{QS}}   \\ \hline
\multicolumn{1}{|c|}{R01}              & \multicolumn{1}{c|}{0.0408}       & 0.1661       &  \multicolumn{1}{|c|}{0.0019} \\ \hline
\multicolumn{1}{|c|}{R02}              & \multicolumn{1}{c|}{0.1102}       & 0.4125       &  \multicolumn{1}{|c|}{0.0039} \\ \hline
\multicolumn{1}{|c|}{R03}              & \multicolumn{1}{c|}{0.5729}       & 1.7662       &  \multicolumn{1}{|c|}{0.0160} \\ \hline
\multicolumn{1}{|c|}{R04}              & \multicolumn{1}{c|}{1.0965}       & 2.7951       &  \multicolumn{1}{|c|}{0.0310} \\ \hline
\multicolumn{1}{|c|}{R05}              & \multicolumn{1}{c|}{2.2189}       & 4.2298       &  \multicolumn{1}{|c|}{0.1300} \\ \hline
\multicolumn{1}{|c|}{R06}              & \multicolumn{1}{c|}{2.8484}       & 4.8205       &  \multicolumn{1}{|c|}{0.2500} \\ \hline
\multicolumn{1}{|c|}{No compression}   & \multicolumn{1}{c|}{28.7465}       & 28.7465      &  \multicolumn{1}{|c|}{1.0000} \\ \hline
\end{tabular}
\end{table}

We compressed and segmented each PC, comparing the segmentation results for the decompressed PC against the segmentation results for the uncompressed PC.  
Hence, the distortion is our metric comparing the segmentation with and without the PC undergoing lossy compression.
The rate is calculated as throughput in MB/sec, assuming a capture rate of 10 frames per second. 
We arbitrarily used $\alpha = 1$ in our experiments, but any other value could be used.  

\begin{figure}[t]
\centering
\includegraphics[width=1\linewidth]{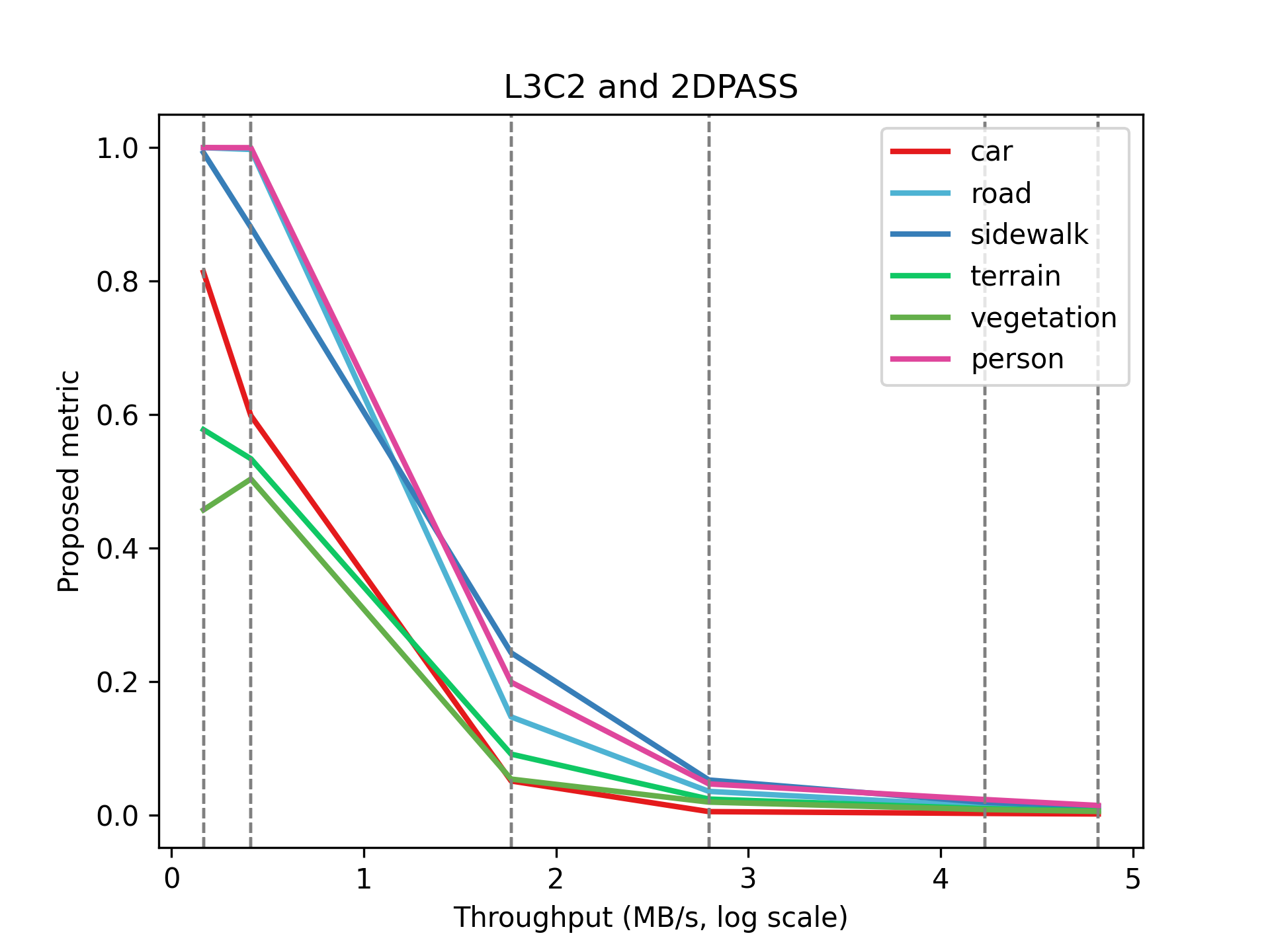}
\caption{The proposed metric results for the L3C2 compression and 2DPASS segmentation algorithm.}
\label{fig:l3c2_2dpass}
\end{figure}

\begin{figure}[t]
\centering
\includegraphics[width=1\linewidth]{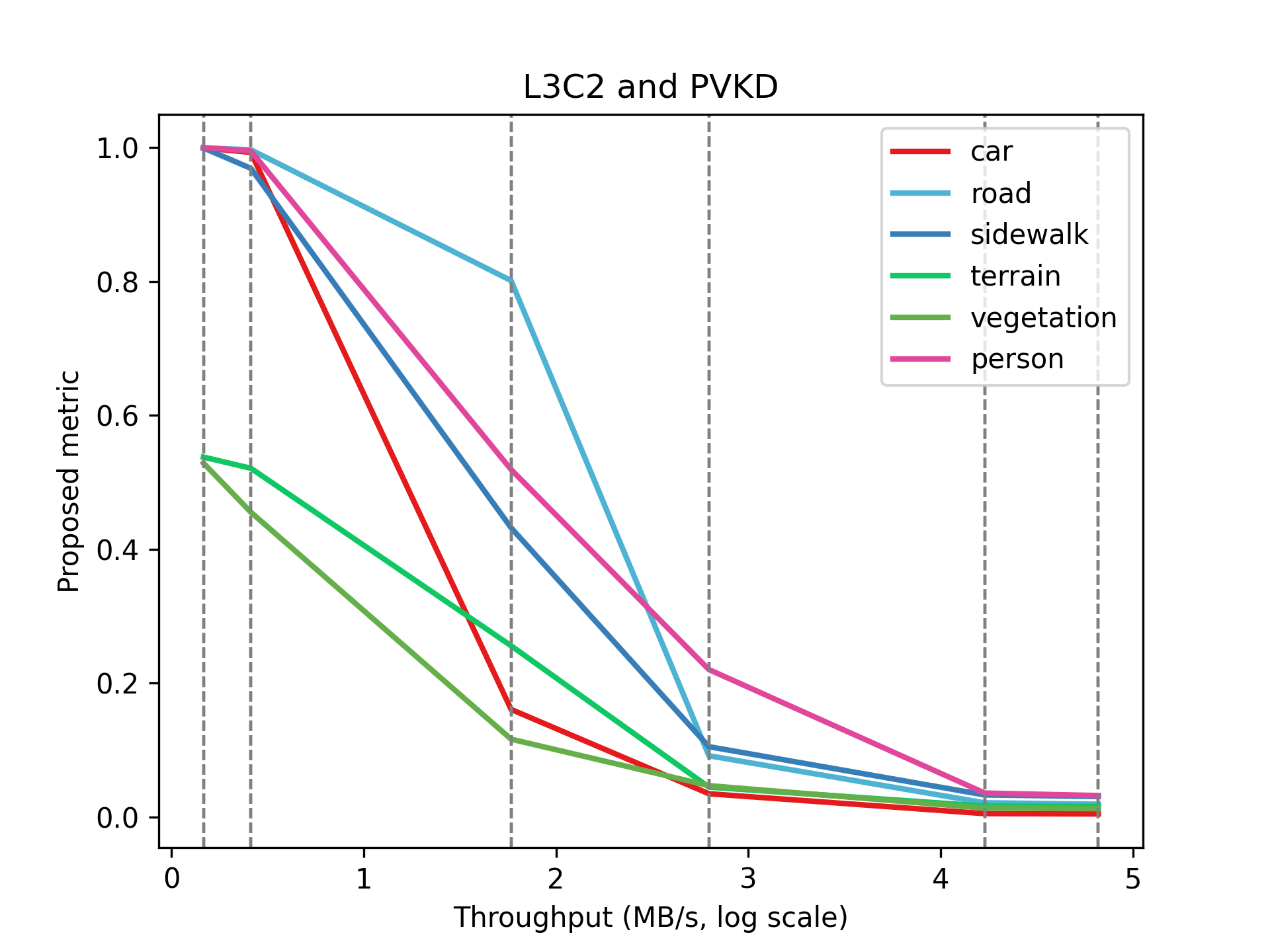}
\caption{The proposed metric results for the L3C2 compression and PVKD segmentation algorithm.}
\label{fig:l3c2_pvkd}
\end{figure}

\begin{figure}[t]
\centering
\includegraphics[width=1\linewidth]{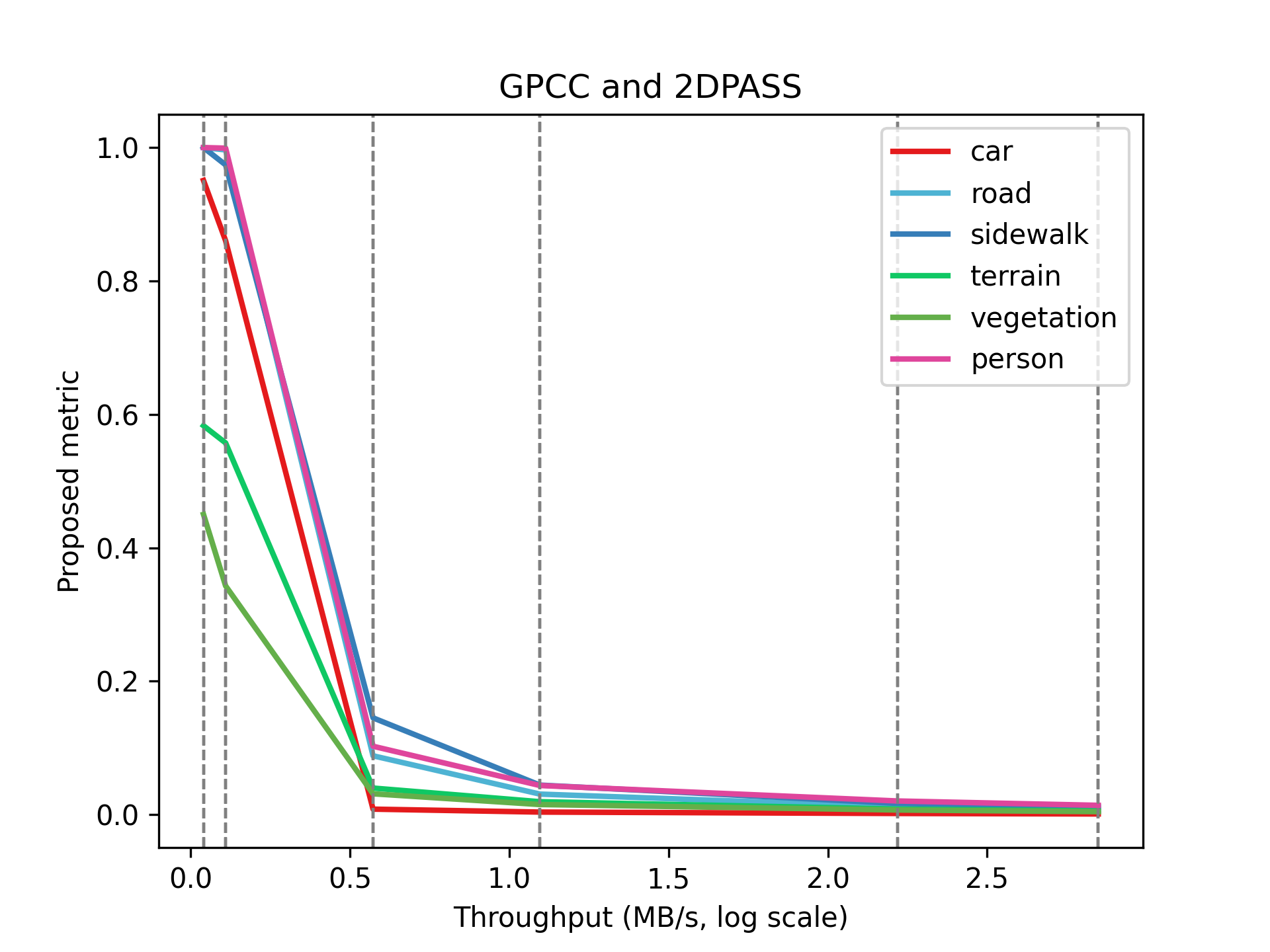}
\caption{The proposed metric results for the G-PCC compression and 2DPASS segmentation algorithm.}
\label{fig:gpcc_2dpass}
\end{figure}

\begin{figure}[t]
\centering
\includegraphics[width=1\linewidth]{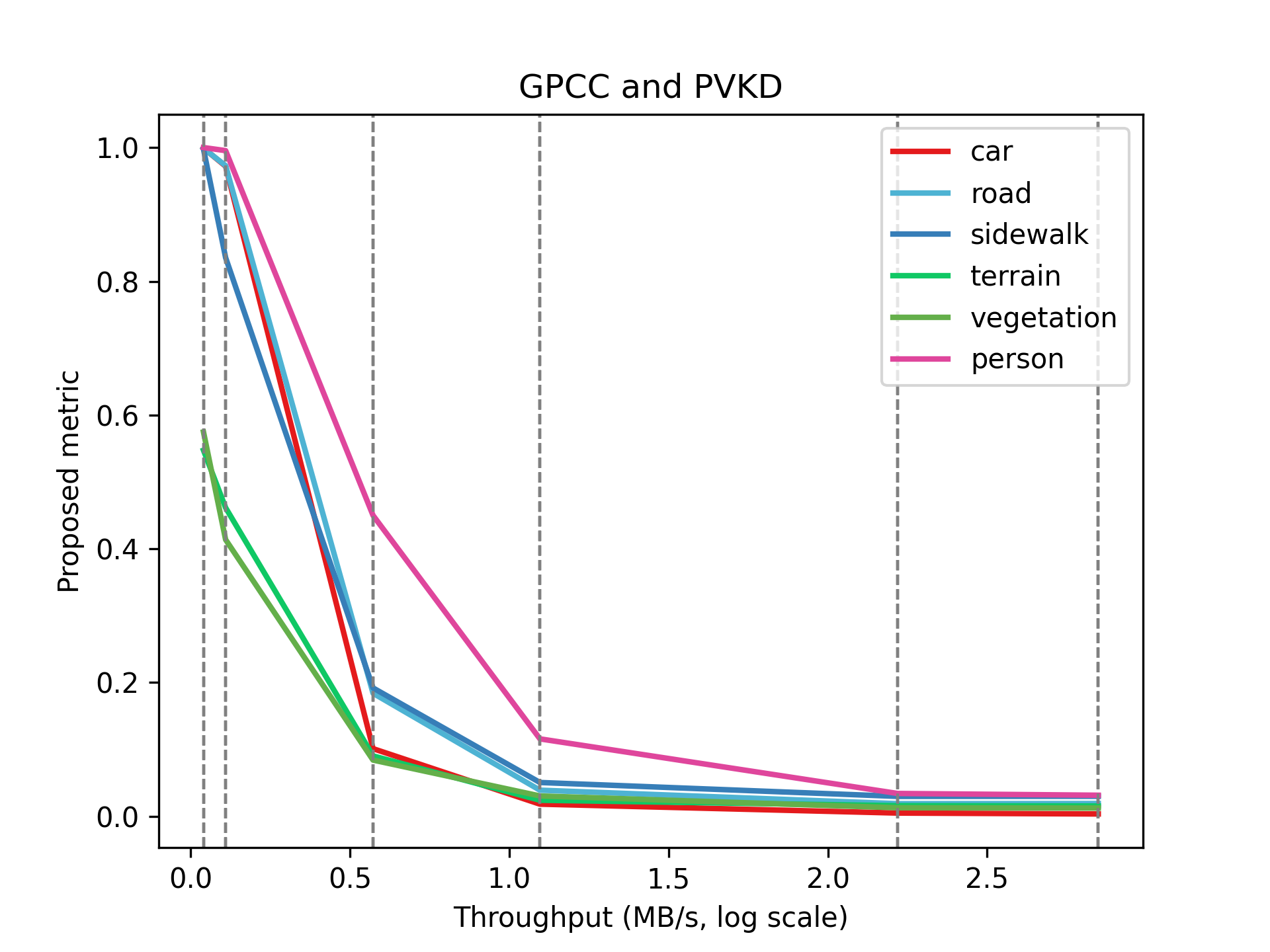}
\caption{The proposed metric results for the G-PCC compression and PVKD segmentation algorithm.}
\label{fig:gpcc_pvkd}
\end{figure}

\begin{figure}[h]
\centering
\includegraphics[width=1\linewidth]{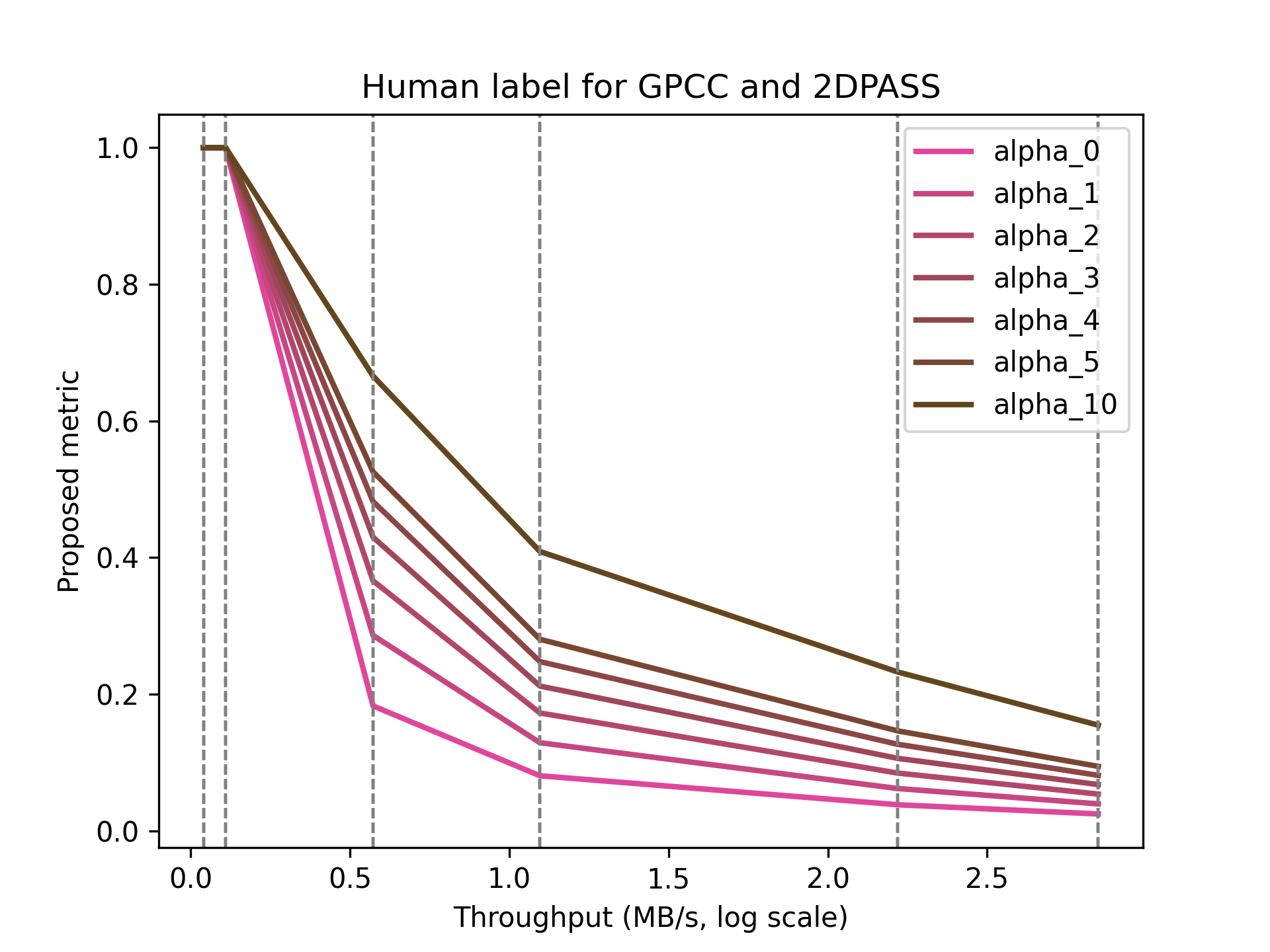}
\caption{The proposed metric results for the Human label, G-PCC compression and 2DPASS segmentation algorithm for different alpha values.}
\label{fig:gpcc_2dpass_alphas}
\end{figure}

Values of QS and the average resulting bit-rates, across the dataset, are shown in Table \ref{table:throughput_qs}.
The rate-distortion (RD) results for each coder-segmenter combination are shown in Figs.
\ref{fig:l3c2_2dpass}, \ref{fig:l3c2_pvkd}, \ref{fig:gpcc_2dpass}, and \ref{fig:gpcc_pvkd}, for a few popular segmentation labels.
These plots are for the averages across the dataset.
Bit rates associated to the {\em R0n} rate parameters are indicated with vertical dashed lines.
The curve behavior for the human label associated with several values of $\alpha$ is shown in Fig. \ref{fig:gpcc_2dpass_alphas}.

\begin{figure}[!t]
\centering
\includegraphics[width=0.9\linewidth]{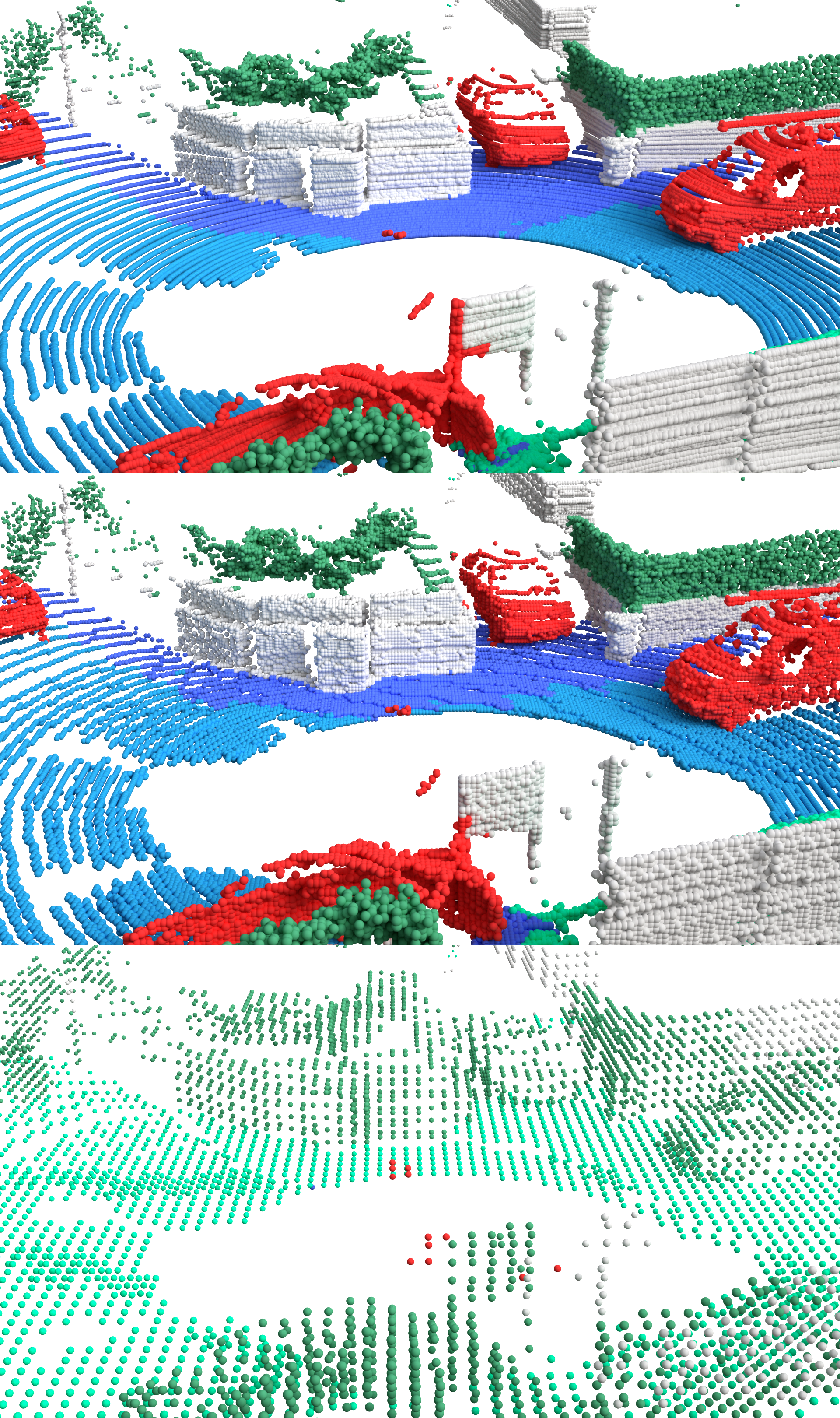}
\caption{Frame from Sequence 11 using G-PCC and 2DPASS. From top to bottom: uncompressed, R03 and R02.}
\label{fig:scan_11}
\end{figure}

\begin{figure}[!t]
\centering
\includegraphics[width=0.9\linewidth]{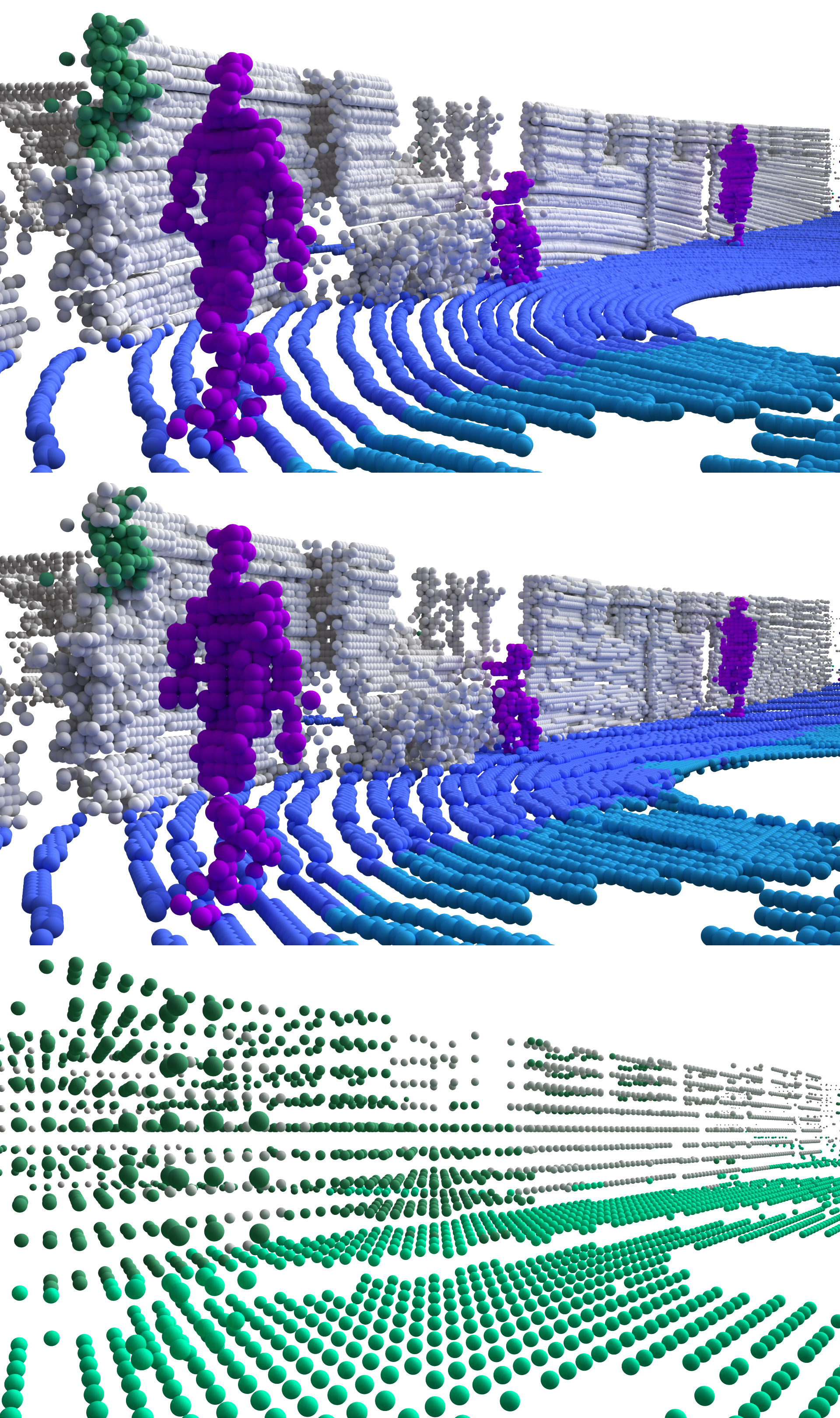}
\caption{Frame from Sequence 18 using G-PCC and 2DPASS. From top to bottom: uncompressed, R03 and R02.}
\label{fig:scan_18}
\end{figure}

In the RD plots, both the G-PCC and L3C2 compressors demonstrate similar results, with the latter requiring higher bit-rates, as expected. Both 2DPASS and PVKD segmentation algorithms exhibit similar performance, but the 2DPASS seems to be slightly more robust against compression artifacts. 
For low rates, distortion is high because the segmentation errors (compared to uncompressed PCs) are too pronounced. 
Examples of such degradation are shown in Figures \ref{fig:scan_11} and \ref{fig:scan_18}, for G-PCC and 2DPASS. They show views of segmented results on uncompressed and compressed point clouds using {\em R03} and {\em R02}. Note that the quality drop from {\em R03} to {\em R02} caused most labels to disappear, thus largely increasing the distortion.



From the RD plots, we conclude that rates {\em R05} and above yield segmentation nearly as good as the original, but perhaps {\em R04} should be good enough for most applications. 
There is a large distortion drop from {\em R02} to {\em R03}, while {\em R01} and {\em R02} are likely not acceptable. 
Perhaps {\em R03} (0.57 MB/s for G-PCC or 1.77 MB/s for L3C2) should be acceptable settings, except for the L3C2 and PVKD combination wherein R04 (2.8 MB/s) may possibly be an acceptable minimum.

Although we just used the Semantic KITTI dataset, we expect other datasets, with similar scan resolution, to yield similar results.



\section{Conclusions}

We explored the impact of point cloud compression on semantic segmentation algorithms, focusing on the G-PCC and L3C2 compressors and the 2DPASS and PVKD segmentation algorithms. Using a new and suitable metric, we evaluated how different compression rates affect segmentation quality within the Semantic KITTI dataset. The results indicate that maintaining high segmentation quality requires throughput values of at least 0.6 MB/s for G-PCC/2DPASS and 2.8 MB/s for L3C2/PVKD. This information is important for ensuring efficiency in autonomous systems, specially in a scenario where data is remotely processed.

The significant drop in segmentation quality, observed when transitioning from {\em R02} to {\em R04} compression settings, may be caused by geometry quantization. The reduced point density at {\em R02} erases important features necessary for accurate labeling. 
Future work may look into a more segmentation-friendly geometry compression. We also need to incorporate some higher-level alarms for when human objects completely appear or disappear. 

\nocite{literature_ref_1}
\nocite{literature_ref_2}
\nocite{literature_ref_3}

\clearpage

\bibliographystyle{IEEEtran}
\bibliography{bibliography.bib}

\begin{thebibliography}{10}
\providecommand{\url}[1]{#1}
\csname url@samestyle\endcsname
\providecommand{\newblock}{\relax}
\providecommand{\bibinfo}[2]{#2}
\providecommand{\BIBentrySTDinterwordspacing}{\spaceskip=0pt\relax}
\providecommand{\BIBentryALTinterwordstretchfactor}{4}
\providecommand{\BIBentryALTinterwordspacing}{\spaceskip=\fontdimen2\font plus
\BIBentryALTinterwordstretchfactor\fontdimen3\font minus \fontdimen4\font\relax}
\providecommand{\BIBforeignlanguage}[2]{{%
\expandafter\ifx\csname l@#1\endcsname\relax
\typeout{** WARNING: IEEEtran.bst: No hyphenation pattern has been}%
\typeout{** loaded for the language `#1'. Using the pattern for}%
\typeout{** the default language instead.}%
\else
\language=\csname l@#1\endcsname
\fi
#2}}
\providecommand{\BIBdecl}{\relax}
\BIBdecl

\bibitem{lidar_autonomous_vehicles}
Y.~Li and J.~Ibanez-Guzman, ``{LiDAR} for autonomous driving: The principles, challenges, and trends for automotive {LiDAR} and perception systems,'' \emph{IEEE Signal Processing Magazine}, vol.~37, no.~4, pp. 50--61, July 2020.

\bibitem{pc_processing_autonomous_driving}
R.~Abbasi, A.~K. Bashir, H.~J. Alyamani, F.~Amin, J.~Doh, and J.~Chen, ``{LiDAR} point cloud compression, processing and learning for autonomous driving,'' \emph{IEEE Transactions on Intelligent Transportation Systems}, vol.~24, no.~1, pp. 962--979, January 2023.

\bibitem{8571288}
S.~Schwarz, M.~Preda, V.~Baroncini, M.~Budagavi, P.~Cesar, P.~A. Chou, R.~A. Cohen, M.~Krivokuća, S.~Lasserre, Z.~Li, J.~Llach, K.~Mammou, R.~Mekuria, O.~Nakagami, E.~Siahaan, A.~Tabatabai, A.~M. Tourapis, and V.~Zakharchenko, ``Emerging {MPEG} standards for point cloud compression,'' \emph{IEEE Journal on Emerging and Selected Topics in Circuits and Systems}, vol.~9, no.~1, pp. 133--148, March 2019.

\bibitem{lidar_semantic_segmentation_survey}
B.~Gao, Y.~Pan, C.~Li, S.~Geng, and H.~Zhao, ``Are we hungry for {3D LiDAR} data for semantic segmentation? a survey of datasets and methods,'' \emph{IEEE Transactions on Intelligent Transportation Systems}, vol.~23, no.~7, pp. 6063--6081, July 2022.

\bibitem{semantic_kitti}
J.~Behley, M.~Garbade, A.~Milioto, J.~Quenzel, S.~Behnke, C.~Stachniss, and J.~Gall, ``{SemanticKITTI}: A dataset for semantic scene understanding of {LiDAR} sequences,'' in \emph{2019 IEEE/CVF International Conference on Computer Vision (ICCV)}, Seoul, South Korea, February 2019.

\bibitem{yan20222dpass}
X.~Yan, J.~Gao, C.~Zheng, C.~Zheng, R.~Zhang, S.~Cui, and Z.~Li, ``{2DPASS}: {2D} priors assisted semantic segmentation on {LiDAR} point clouds,'' in \emph{European Conference on Computer Vision}, Tel Aviv, Israel, July 2022.

\bibitem{pvkd}
Y.~Hou, X.~Zhu, Y.~Ma, C.~C. Loy, and Y.~Li, ``Point-to-voxel knowledge distillation for {LiDAR} semantic segmentation,'' in \emph{IEEE Conference on Computer Vision and Pattern Recognition}, New Orleans, United States, June 2022.

\bibitem{gpcc}
D.~Graziosi, O.~Nakagami, S.~Kuma, A.~Zaghetto, T.~Suzuki, and A.~Tabatabai, ``An overview of ongoing point cloud compression standardization activities: video-based ({V-PCC}) and geometry-based ({G-PCC}),'' \emph{APSIPA Transactions on Signal and Information Processing}, vol.~9, p. e13, April 2020.

\bibitem{l3c2}
S.~Lasserre and J.~Taquet, ``A point cloud codec for {LiDAR} data with very low complexity and latency.'' m56477. ISO/IEC JTC 1/SC 29/WG 7., 2021.

\bibitem{Queiroz2016}
R.~L. de~Queiroz and P.~A. Chou, ``Compression of {3D} point clouds using a region-adaptive hierarchical transform,'' \emph{{IEEE} Trans. Image Process.}, vol.~25, no.~8, pp. 3947--3956, August 2016.

\bibitem{literature_ref_1}
N.~A.~B. Martins, L.~A. d.~S. Cruz, and F.~Lopes, ``Impact of lidar point cloud compression on {3D} object detection evaluated on the {KITTI} dataset,'' \emph{J. Image Video Process.}, vol. 2024, no.~1, June 2024.

\bibitem{literature_ref_2}
L.~Garrote, J.~Perdiz, L.~A. da~Silva~Cruz, and U.~J. Nunes, ``Point cloud compression: Impact on object detection in outdoor contexts,'' \emph{Sensors}, vol.~22, no.~15, August 2022.

\bibitem{literature_ref_3}
M.~Roque and L.~A. da~Silva~Cruz, ``An empirical study of the effect of point cloud compression on the performance of segmentation,'' in \emph{2019 8th European Workshop on Visual Information Processing (EUVIP)}, January 2019.

\bibitem{ctc_ref}
``{Common test conditions for point cloud compression},'' ISO/IEC JTC1/SC29 Joint WG11/WG1 (MPEG/JPEG), Gothenburg, SE, Approved WG 11 doc. N18883, July 2019.

\end{thebibliography}

\end{document}